\documentclass[reprint,amsmath,amssymb,floatfix,twocolumn,superscriptaddress,aps,pre]{revtex4-2}
\usepackage{hyperref}
\usepackage{graphicx}
\usepackage{dcolumn}% Align table columns on decimal point
\usepackage{threeparttable}
\usepackage{bm, mathtools}% bold math
\usepackage{color}
\usepackage{xcolor}
\hypersetup{
    colorlinks=true,
    linkcolor=blue,
    citecolor=blue,   
   urlcolor=blue,
   }

\newcommand {\pwisein}{\left\{ \begin{array}{ll}}
\newcommand {\pwiseout}{\end{array}\right.}

\newcommand {\dl}{\delta}
\newcommand {\D}{\Delta}

\newcommand {\jtilde}{\langle \tilde{j} \rangle}

 \graphicspath{{figures_new/}}
 
\begin{document}

\title{Asymmetry-controlled resonant transport in a Brownian flashing ratchet}

\author{Sudipta Mandal}
\email{ph16049@iisermohali.ac.in}
\email{sudiptam@tifrh.res.in}
\altaffiliation[Present Address:]{Tata institute of fundamental
  research,Hyderabad
36/P, Gopanpally Village, Serilingampally Mandal, Ranga Reddy
District, Hyderabad, Telangana 500046}
\affiliation{Department of Physical Science, Indian Institute of
  Science Education and Research Mohali,Sector 81, S. A. S
  Nagar,Manauli PO 140306,India}

\author{Dipanjan Chakraborty}
\email{chakraborty@iisermohali.ac.in}
\affiliation{Department of Physical Science, Indian Institute of
  Science Education and Research Mohali,Sector 81, S. A. S
  Nagar,Manauli PO 140306,India}

\author{Debasish Chaudhuri}
\email{debc@iopb.ac.in}
\affiliation{Institute of Physics, P.O: Sainik School,
  Bhubaneswar-751005, Odisha, India}

\date{\today}

  \begin{abstract}
We investigate directed transport in a one-dimensional Brownian flashing
ratchet with a piecewise-linear asymmetric periodic potential. Numerical
solutions of the Fokker--Planck equation and Brownian dynamics simulations
reveal a nonmonotonic dependence of the stationary current on the switching
frequency, with a resonant maximum whose position depends on the potential
asymmetry $\delta$ and barrier height $\Delta$. For moderate asymmetry,
$|\delta|<0.5$, the current obeys a scaling form that separates the dependence
on the potential parameters from a common frequency dependence, resulting in
a data collapse upon appropriate scaling of the current and frequency. The
current amplitude varies linearly with $\delta$, while the resonance
frequency follows
$\nu(\delta,\Delta)=\nu_0(\Delta)/ [1-b_0\, \delta^2]$.
We interpret the resulting $(1-\delta^2)^{-1}$ scaling in terms of coupled
relaxation along the two branches of the asymmetric potential, which
provides a physical basis for the observed dependence of the resonant
frequency on the potential asymmetry.
\end{abstract}

%For fun, see PACS list at  https://www.aip.org/publishing/pacs/pacs-2010-regular-edition
%\pacs{47.15.-x}

\maketitle

\section{Introduction}

Directed transport in systems without a net bias is a characteristic
signature of nonequilibrium statistical physics. A paradigmatic example is
the flashing ratchet, in which thermal fluctuations are rectified by
periodically or stochastically switching an asymmetric potential
\cite{
Bug1987Shaking,Doering1994Nonequilibrium,Astumian1994, 
Ajdari1992,Julicher1995,julicher1997modeling,Prost1994,Astumina1997, astumian2002brownian,
  hanggi2009artificial}. Since the early studies of fluctuation-driven transport, the
flashing-ratchet mechanism has been investigated extensively in the context
of molecular motors and Brownian transport, and has also been realized
experimentally in colloidal systems \cite{Rousselet1994,Faucheux1995,
 Harmer2001Brownian, reimann2002brownian,hanggi2009artificial, Marquet2002,Germs2013,
Tierno2010,Tierno2012, Gerayeli2021, Evers2013,Han2023,Han2025}.

The directed current in a flashing ratchet depends on the competition
between the switching timescale and the intrinsic relaxation times of the
particle in the potential. Consequently, the current is generally
non-monotonic as a function of the switching frequency, exhibiting a
maximum at an intermediate frequency. Such frequency-dependent transport
has been studied for a variety of ratchet potentials, including
piecewise-linear and sawtooth potentials
\cite{Makhnovskii2004,Rozenbaum2012,Rozenbaum2016HighTemperature,
Shapochkina2019Relaxation,Saakian2018}. The connection between the
switching frequency and characteristic relaxation times has also been
investigated in detail, both for single-particle and interacting ratchet
systems \cite{Ajdari1992, Prost1994, Julicher1995, julicher1997modeling, hanggi2009artificial, Chakraborty2015stochastic, Khali2020structure}.

Despite this extensive literature, the dependence of the characteristic
resonance frequency on the geometric asymmetry of a flashing potential is
less transparent. This question is particularly natural for a
piecewise-linear ratchet, whose two branches have lengths $s_1$ and $s_2$
with $s_1+s_2=\lambda$. Existing analytical treatments of flashing
 ratchets have demonstrated that the finite-frequency current is
controlled by the spectrum of the corresponding Fokker--Planck operator,
which becomes nontrivial even for piecewise-linear potentials
\cite{Klosek1999,Makhnovskii2004,Saakian2018}. Simple estimates based on
independent relaxation times of the two branches therefore need not capture
the relaxation mode responsible for the resonance.

In this work, we study a minimal one-dimensional flashing ratchet with a
piecewise-linear asymmetric potential and systematically investigate how
its resonance frequency depends on the potential geometry and barrier
height $\Delta$. We characterize the asymmetry by the dimensionless parameter
$\delta=(s_2-s_1)/(s_1+s_2)$, which changes sign upon inversion of the
ratchet while leaving the resonance frequency unchanged. We determine the
stationary current from the Fokker--Planck equation and independently from
Brownian-dynamics simulations. We find that the current exhibits a
well-defined maximum at an intermediate switching frequency and that the
corresponding resonance frequency increases systematically with the
geometric asymmetry. Over a broad range of parameters, the dependence is
accurately described by an inverse-quadratic form,
$\nu(\delta,\Delta)\simeq\nu_0(\Delta)/[1-b_0\, \delta^2]$, with
$b_0$ close to unity.

We further show that this behavior cannot be captured by simply adding the
characteristic ``on'' localization and ``off'' diffusion times of the two
branches. Instead, the observed dependence can be rationalized by viewing
the resonance as a collective relaxation process involving both sides of
the asymmetric potential. The resulting scaling connects the resonance
frequency to the product of the two branch lengths and provides a simple
geometrical interpretation of the numerical results.

\section{Model}

We consider an overdamped Brownian particle moving in a one-dimensional
periodic asymmetric potential that is switched randomly between an ON state
and an OFF state. The particle obeys

\begin{equation}
\dot{x}=-\mu \frac{\partial U(x,t)}{\partial x}
+\sqrt{2D_0}\,\xi(t),
\end{equation}
where $\mu$ is the mobility, $D_0=\mu k_{\rm B}T$ is the bare diffusion
coefficient, and $\xi(t)$ is a zero-mean unit-variance Gaussian white noise.
The time-dependent potential is written as

\begin{equation}
U(x,t)=U_0\,\sigma (t)\,u(x),
\end{equation}
where $U_0$ is the barrier height and $\sigma (t)=1$ when the ratchet is ON and
$\sigma (t)=0$ when it is OFF.

The spatial part of the potential is periodic with period $\lambda$,
$u(x+\lambda)=u(x)$. Within one period, we choose a piecewise-linear
profile with a minimum at $x=s_1$ and a maximum at the cell boundaries.
Equivalently, the two branches of the potential have lengths $s_1$ and
$s_2$, with

\begin{equation}
s_1+s_2=\lambda.
\end{equation}

It is convenient to characterize the asymmetry by

\begin{equation}
\delta=\frac{s_2-s_1}{s_1+s_2},
\end{equation}

so that

\begin{equation}
s_1=\frac{\lambda}{2}(1-\delta), \qquad s_2=\frac{\lambda}{2}(1+\delta).
\end{equation}

With this definition, $\delta=0$ corresponds to a symmetric potential,
whereas changing $\delta$ to $-\delta$ simply interchanges the two
branches. The barrier height is measured in thermal units through

\begin{equation}
\Delta=\frac{U_0}{k_{\rm B}T}.
\end{equation}

The slopes of the two branches are therefore fixed by the same barrier
height and are inversely proportional to their lengths. This parametrization
allows the period $\lambda$, asymmetry $\delta$, and barrier height
$\Delta$ to be varied independently.

\subsection{Stochastic flashing}

The ratchet is switched between the ON and OFF states according to a
symmetric two-state Markov process. The transition rates are taken to be
equal,

\begin{equation}
0 \xrightleftharpoons[r]{r} 1,
\end{equation}
where $r$ is the switching rate. Thus the mean residence time in either
state is $1/r$, and the characteristic diffusion time over one spatial
period is

\begin{equation}
\tau_D=\frac{\lambda^2}{D_0}.
\end{equation}
The dimensionless switching frequency used below is

\begin{equation}
f=r\tau_D.
\end{equation}

The symmetric switching protocol is important here because it introduces no
additional temporal bias: directed transport arises solely from the
spatial asymmetry of the potential combined with thermal fluctuations.

After measuring length in units of $\lambda$ and time in units of
$\tau_D$, the dynamics is controlled by only two independent parameters,
the asymmetry $\delta$ and the barrier height $\Delta$, in addition to the
dimensionless switching frequency $f$.

\begin{figure}[!t]
 \includegraphics[width=1\linewidth]{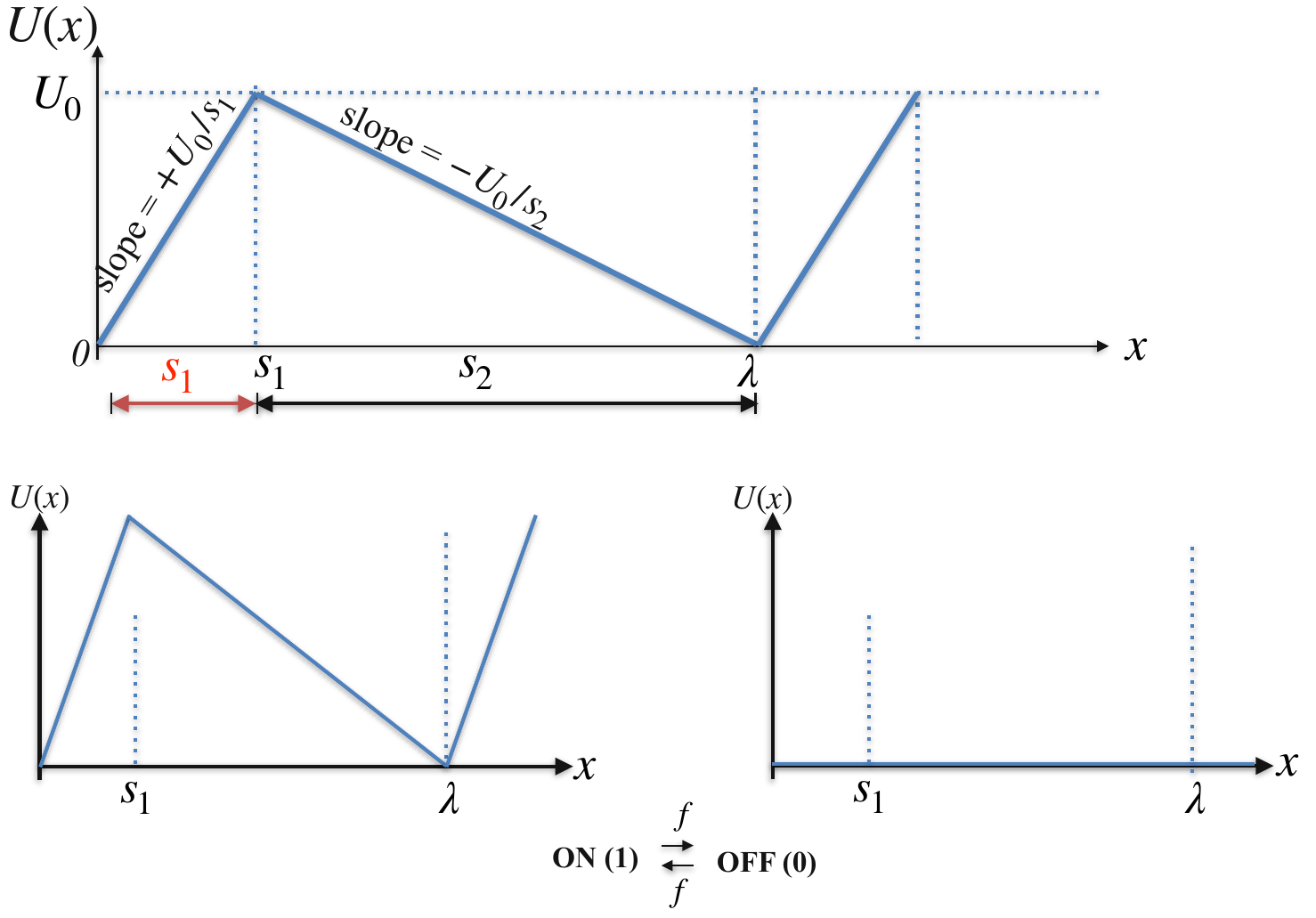} %{ratchet.png}
 \caption{Schematic of the flashing linear ratchet model. The top
   panel shows one spatial period of the asymmetric piecewise-linear
   potential in the ON state, with period $\lambda=s_1+s_2$, barrier
   height $U_0$, and minimum at $x=s_1$. The two linear branches have
   lengths $s_1$ and $s_2$ and slopes $-U_0/s_1$ and $U_0/s_2$,
   respectively. The asymmetry is characterized by
   $\delta=(s_2-s_1)/(s_1+s_2)$. The lower panels show the periodic
   potential in the ON state ($s=1$) and the flat potential in the OFF
   state ($s=0$). The potential switches stochastically between the
   two states with equal transition rates $f$.}
   \label{fig:potential}
\end{figure}

\subsection{Fokker--Planck description}

To calculate the stationary current, we introduce the probability densities
$P_{\rm on}(x,t)$ and $P_{\rm off}(x,t)$ conditioned on the ratchet being
in the ON and OFF states, respectively. They obey the coupled equations

\begin{equation}
\partial_t P_{\rm on} =-\partial_x J_{\rm on} -fP_{\rm on} +fP_{\rm off},
\end{equation}

\begin{equation}
\partial_t P_{\rm off} = -\partial_x J_{\rm off} +fP_{\rm on} -fP_{\rm off},
\end{equation}

with probability currents

\begin{equation}
J_{\rm on} = -\mu U'(x)P_{\rm on} -D_0\partial_x P_{\rm on},
\end{equation}

\begin{equation}
J_{\rm off} = -D_0\partial_x P_{\rm off}.
\end{equation}

The probability densities satisfy periodic boundary conditions over one
unit cell and are normalized according to

\begin{equation}
\int_0^\lambda \left[P_{\rm on}(x,t)+P_{\rm off}(x,t)\right]\,dx=1.
\end{equation}

In the stationary state, the total current
$J(x)=J_{\rm on}(x)+J_{\rm off}(x)$ is independent of $x$. The mean
particle current is therefore obtained directly from this constant
stationary flux.

\subsection{Numerical methods}

We obtain the stationary solution of the coupled Fokker--Planck equations
numerically over one spatial period and determine the corresponding
stationary current from the total probability flux (see Appendix-\ref{sec_appA}). As an independent
check, we integrate the overdamped Langevin equation using the
Euler--Maruyama scheme. The unit of length and time are set by the spatial period $\lambda$ and
$\tau_D= \lambda^2/D_0$, respectively. The numerical time step is chosen as
$\Delta t=10^{-3}\tau_D$. The flashing protocol is implemented as a
two-state Markov process with transition rate $f$. The stationary current
is measured from the long-time displacement of the particles.

\section{Results}

We first consider the stationary particle current obtained from the
Fokker--Planck equation and compare it with Brownian-dynamics simulations.
 The two numerical approaches give consistent
results throughout the parameter range considered.

\subsection{Frequency dependence of the particle current}

\begin{figure}[!t]
  \centering
  \includegraphics[width=\linewidth]{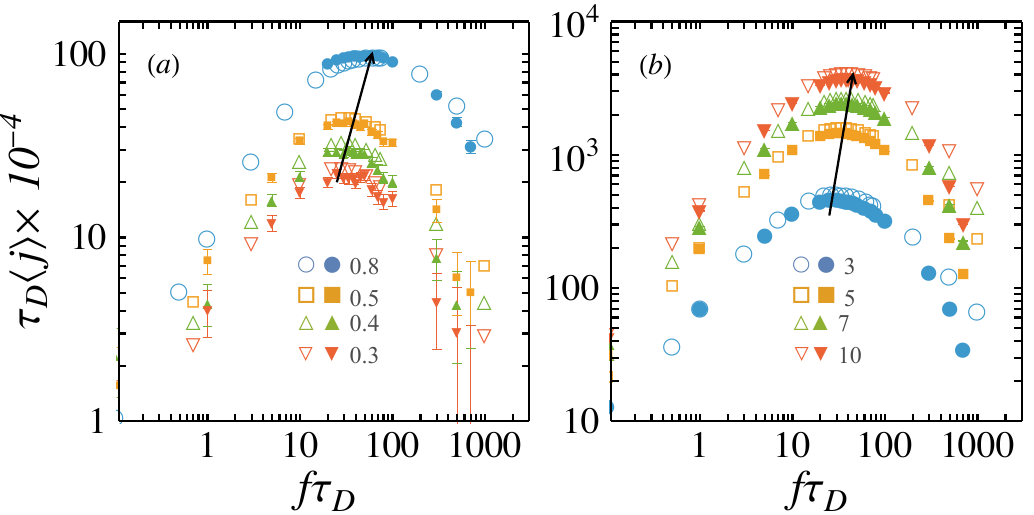}    
  \caption{ $\bf{(a)}$ Particle current versus driving frequency for
    different asymmetry parameters $\delta$ at $\Delta=1$. $\bf{(b)}$
    Particle current versus driving frequency for different potential
    heights $\Delta$ at $\delta=0.3$. Open and filled symbols denote
    Fokker–Planck and Brownian dynamics results, respectively. Arrows
    indicate the shift of the resonance frequency with increasing
    $\delta$ or $\Delta$.  }
  \label{fig:current_raw_data}
\end{figure}

We begin by examining the dependence of the particle current on the
switching frequency. Figure~\ref{fig:current_raw_data}(a) shows the
dimensionless stationary current $\langle\tilde{j}\rangle$ as a function of
$f$ for several values of the asymmetry parameter $\delta$ at fixed
potential strength $\Delta=1$. The current is non-monotonic in the
switching frequency: it increases at low frequencies, reaches a maximum at
an intermediate frequency, and decreases again at high frequencies. The
open and filled symbols, corresponding respectively to the numerical
solution of the Fokker--Planck equation and Brownian-dynamics simulations,
show good agreement.

The position of the maximum defines a characteristic resonance frequency
$\nu$. An important observation is that $\nu$ is not fixed by the diffusion
time alone. As the asymmetry is increased, the position of the maximum
systematically shifts towards higher frequency. The same behavior is
observed when the potential strength is varied, as shown in
Fig.~\ref{fig:current_raw_data}(b): increasing $\Delta$ shifts the maximum
to higher frequency. Thus, the resonance frequency is a function of both
the geometric asymmetry and the potential strength, $\nu=\nu(\delta,\Delta)$.

The non-monotonic behavior has a simple interpretation in terms of the
competition between the switching and relaxation processes. At low
frequency, the particle has sufficient time to relax during each cycle,
but the number of switching events per unit time is small. Consequently,
the current vanishes as $f\rightarrow0$. At high frequency, on the other
hand, the particle distribution cannot follow the rapidly changing
potential, and the directed current again decreases. The detailed
asymptotic behavior of these two regimes is discussed in
Appendix~\ref{app:high_low_frequency}.

\subsection{Amplitude and frequency scaling}

The numerical results suggest that, over an appropriate range of
parameters, the current can be separated into an amplitude that depends on
the potential parameters and a function describing its frequency
dependence. We therefore write

\begin{equation}
\langle\tilde{j}\rangle = \widetilde{J}(\delta,\Delta)\,g(\nu,f),
\label{eq:current_scaling}
\end{equation}
where $\widetilde{J}(\delta,\Delta)$ denotes the amplitude of the current
and $\nu$ is the resonance frequency.

  \begin{figure}[!t]
  \centering
  \includegraphics[width=\linewidth]{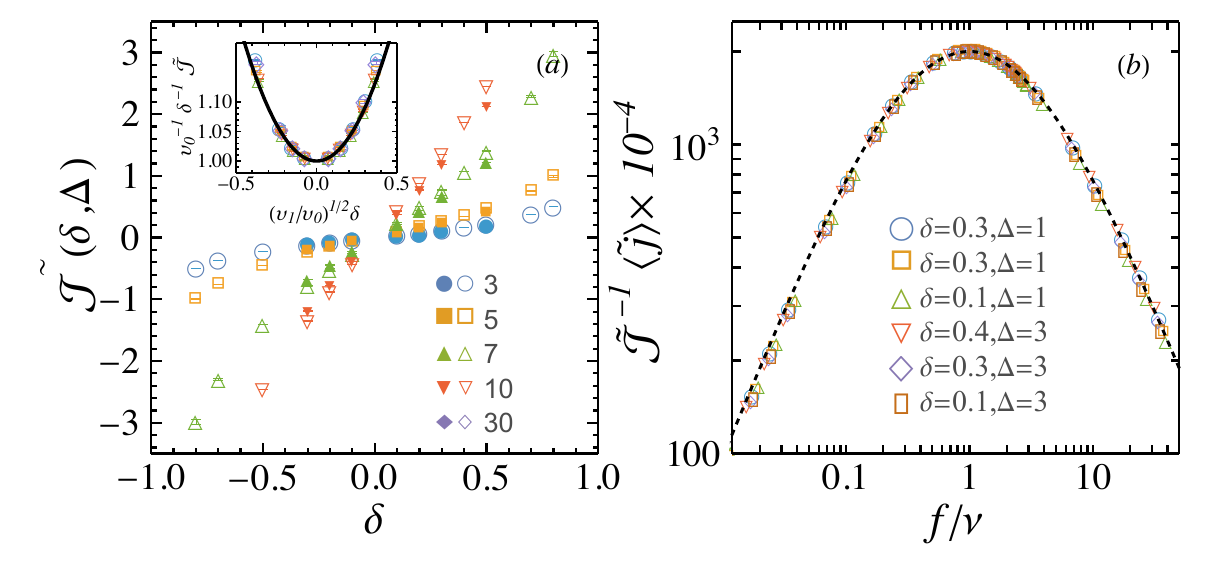} 
  \caption{\textbf{(a)}: Estimated current amplitude
    $\tilde{\mathcal{J}}(\delta,\Delta)$ from fits to the
    Fokker–Planck (open symbols) and Brownian-dynamics (filled
    symbols) results for different $\Delta$ values. Inset: data
    collapse of $\tilde{\mathcal{J}}/v_0 \delta$ versus $\delta$, showing the
    expected parabolic dependence $1+(v_1/v_0)\delta^2$.
    \textbf{(b)}: Plot of $\langle\tilde{j}\rangle/\widetilde{J}$
    versus the scaled frequency $f/\nu$ demonstrating data
    collapse. The dashed line corresponds to Eq.~\eqref{eq:g_function}
    with $c=3$.  }
 \label{fig:current_dala_collapse}
\end{figure}

We first examine the dependence of the amplitude on the asymmetry. Since
changing $\delta$ to $-\delta$ corresponds to spatial inversion of the
ratchet, the direction of the current changes while its magnitude remains
unchanged. Consequently,
$
\widetilde{J}(-\delta,\Delta) = -\widetilde{J}(\delta,\Delta),
$
so that the amplitude must be an odd function of $\delta$. For weak
asymmetry, it can therefore be expanded as

\begin{equation}
\widetilde{J}(\delta,\Delta)
=
v_0(\Delta)\delta
+
v_1(\Delta)\delta^3+\cdots.
\label{eq:amplitude_expansion}
\end{equation}

The amplitude obtained from the numerical current is shown in
Fig.~\ref{fig:current_dala_collapse}(a) for different values of the
potential strength. For small $|\delta|$, the dependence is
predominantly linear in $\delta$, whereas deviations from linearity become
increasingly important for relatively larger asymmetry. As we show in the Appendix-\ref{sec:current_on_Delta}, 
both $v_0$ and $v_1$ increase with $\Delta$.

We next ask whether the remaining frequency dependence can be described by
a single characteristic frequency $\nu$. If the factorization in
Eq.~\eqref{eq:current_scaling} is valid, the scaled current
$\langle\tilde{j}\rangle/\widetilde{J}$ should depend only on the reduced
frequency $f/\nu$. Figure~\ref{fig:current_dala_collapse}(b) shows that this
is indeed approximately the case in the regime of $|\dl| < 0.5$ and $\Delta < 10$. The data obtained from both the
Fokker--Planck equation and Brownian dynamics collapse onto a common curve
when the current is scaled by its amplitude and the
frequency is normalized by the resonance frequency $\nu$. The quality of the collapse deteriorates
for larger asymmetry, consistently with the increasing importance of the
higher-order terms in Eq.~\eqref{eq:amplitude_expansion}.

The two limiting behaviors of the frequency-dependent part can be obtained
from the low- and high-frequency regimes. As discussed in Appendix~\ref{app:high_low_frequency}, the current scales 
linearly with the switching frequency in the slow-switching limit, $\langle\tilde{j}\rangle\sim f$ 
as $f\to0$, whereas in the fast-switching limit it decays as 
$\langle\tilde{j}\rangle\sim f^{-1}$ as $f\to\infty$.

The simplest rational interpolation consistent with both limits and with
a maximum at $f=\nu$ is

\begin{equation}
g(\nu,f) = \frac{\nu f} {\nu^2+c\nu f+f^2}.
\label{eq:g_function}
\end{equation}

The parameter $c$ controls the shape of the frequency response but does not
change the position of its maximum, which remains at $f=\nu$. 
Fitting the collapsed data in Figure~\ref{fig:current_dala_collapse}(b) with Eq.\eqref{eq:g_function} gives
$c\approx 3$. Thus, within the regime $|\delta|<0.5$, the current can be
represented approximately as

\begin{equation}
\langle\tilde{j}\rangle \simeq \widetilde{J}(\delta,\Delta) \frac{\nu f} {\nu^2+3\nu f+f^2}.
\label{eq:current_interpolation}
\end{equation}

\subsection{Resonance frequency and geometric asymmetry}

\begin{figure}[!ht]
 % \centering
  \includegraphics[width=\linewidth]{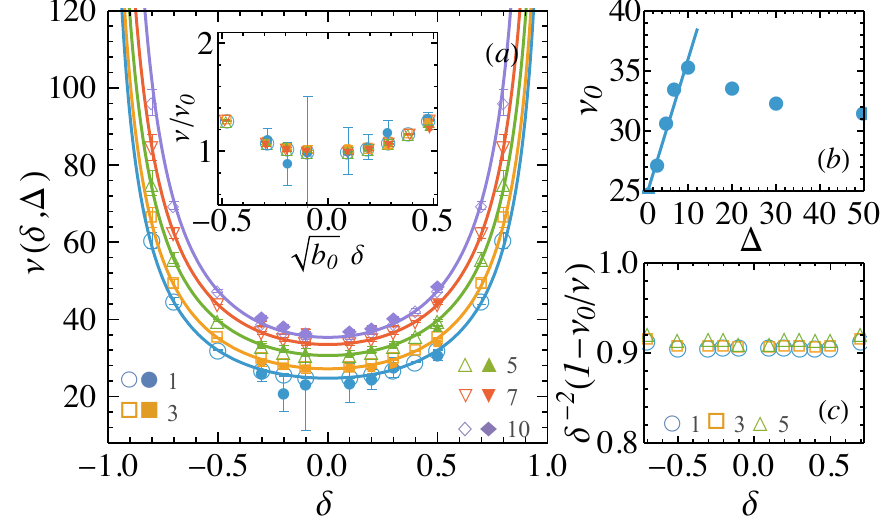}
  \caption{\textbf{(a)} Normalized resonance frequency
    $\nu(\Delta,\delta)$ versus asymmetry $\delta$ for different
    $\Delta$. Open (filled) symbols denote Fokker--Planck
    (Brownian-dynamics) results; solid lines show
    $\nu_0/(1-b_0\delta^2)$. Inset: data collapse of $\nu/\nu_0$
    versus $b_0^{1/2}\delta$ for $|\delta|<0.5$. \textbf{(b)} $\nu_0$
    versus $\Delta$. For $\Delta\leq10$, $\nu_0$ increases
    approximately linearly with $\Delta$ (solid line). \textbf{(c)}
    $b_0=\delta^{-2}(1-\nu_0/\nu)$ versus $\delta$, showing that $b_0$
    remains approximately constant for $|\delta|<0.5$. Symbols
    correspond to the $\Delta$ values in the legend; only
    Fokker--Planck results are shown for clarity.}
 \label{fig:nu_function_of_delta_for_different_Delta}
\end{figure}

Having established that the frequency dependence can be characterized by a
single resonance frequency $\nu$, we now investigate how this frequency
depends on the geometry of the ratchet. Figure~\ref{fig:nu_function_of_delta_for_different_Delta}(a) shows the resonance frequency as a
function of the asymmetry parameter for several values of $\Delta$. The
results from the Fokker--Planck equation and Brownian-dynamics simulations
again agree well. The resonance frequency increases systematically with
$|\delta|$ and the same functional dependence is observed for different
potential strengths.

The invariance of the relaxation dynamics under interchange of the two
branches requires the resonance frequency to be an even function of
$\delta$. The numerical results are accurately described by the
inverse-quadratic form

\begin{equation}
\nu(\delta,\Delta) = \frac{\nu_0(\Delta)} {1-b_0(\Delta)\delta^2} .
\label{eq:nu_delta}
\end{equation}
This form is validated by the inset of Fig.~\ref{fig:nu_function_of_delta_for_different_Delta}(a)
which shows the corresponding data collapse when $\nu/\nu_0$ is plotted against
$b_0^{1/2}\delta$.

The characteristic frequency $\nu_0$ increases with the potential strength,
as shown in Fig.~\ref{fig:nu_function_of_delta_for_different_Delta}(b).
For $\Delta\lesssim10$, $\nu_0$ increases approximately linearly with
$\Delta$,
\begin{equation}
\nu_0(\Delta) \simeq \widetilde{\nu}_0 \left(1+\frac{\Delta}{\Delta_0}\right),
\label{eq:nu0_delta}
\end{equation}
with $\widetilde{\nu}_0\simeq24.76$. At larger barrier heights,
$\nu_0$ decreases to approach a plateau.  The fitted values of $b_0$
remain approximately constant and close to unity over the range where
the inverse-quadratic description applies
[Fig.~\ref{fig:nu_function_of_delta_for_different_Delta}(c)].
Accordingly, we omit the possible $\Delta$ dependence of $b_0$ in what
follows.

Finally, restoring dimensions, the resonance frequency, denoted by the same symbol $\nu$, is given in terms of
the physical parameters of the ratchet by
\begin{equation}
\nu = \frac{D_0}{\lambda^2} \frac{\nu_0(\Delta)} {1-b_0\,\delta^2}.
\label{eq:dimensional_resonance}
\end{equation}
Thus, the spatial period enters through the diffusive timescale
$\lambda^2/D_0$, while the dependence on the shape and strength of the
potential is contained in $\delta$ and $\Delta$.

The numerical observation in Eq.~\eqref{eq:nu_delta} is the central result
of this study. In the following section, we examine whether it can be
understood in terms of the characteristic ON-state localization and
OFF-state diffusion times, and show why a simple sum of these two
timescales fails to reproduce the observed dependence on asymmetry.

\section{Discussion}

\subsection{Geometric origin of the resonance scaling}

The numerical results show that the resonance frequency is well described by
$
\nu(\delta,\Delta) \simeq \frac{\nu_0(\Delta)} {1-b_0\, \delta^2},
$
with $b_0$ close to unity over the range of parameters investigated.
The appearance of the combination $1-\delta^2$ can be related directly to
the geometry of the two branches of the ratchet. 
For the discussion of this section, we retain dimensional quantities. 
In terms of their lengths,
$s_1=\frac{\lambda}{2}(1-\delta)$, $s_2=\frac{\lambda}{2}(1+\delta)$,
so that
\begin{equation}
s_1s_2 = \frac{\lambda^2}{4}(1-\delta^2).
\label{eq:s1s2}
\end{equation}
We now argue that the collective relaxation of the probability distribution
naturally involves this product rather than the sum of the two branch
relaxation times.

As discussed in Appendix~\ref{sec_appA}, the probability distributions on
the two branches form a single coupled Fokker--Planck problem, with the
solutions matched through continuity of both the probability density and
the probability current.
For the ON state, the Fokker--Planck equation on each linear branch can be
written schematically as
\begin{equation}
\partial_t P = D_0\partial_x^2P + D_0F_\alpha\partial_xP,
\end{equation}
where $F_\alpha$ is the constant force on branch $\alpha$. Introducing a coordinate
scaled by the corresponding branch length, $y_\alpha=x/s_\alpha$, gives
\begin{equation}
\partial_t P = \frac{D_0}{s_\alpha^2} \left[ \partial_{y_\alpha}^2P + (F_\alpha s_\alpha)\partial_{y_\alpha}P \right].
\end{equation}

For the present piecewise-linear potential, $F_\alpha s_\alpha$ is determined by the
same dimensionless barrier height on both branches. Thus, apart from a
common dimensionless function of $\Delta$, the characteristic relaxation
rate associated with each spatial sector scales as

$\Gamma_\alpha\sim\frac{D_0}{s_\alpha^2}$.

The important point is that these two relaxation processes are not
independent. The solutions on the two branches form a single eigenmode of
the Fokker--Planck operator and are coupled through the continuity of both
the probability density and the probability current. Consequently, the
relaxation of the complete unit cell cannot be obtained simply by adding
two independent branch relaxation times.

Since the lowest mode involves both spatial
sectors,  its characteristic
scale must be a symmetric function of the two branch lengths. A natural
collective scaling is obtained from the mixed spatial scale
$s_1 s_2$, giving

\begin{equation}
\Gamma_{\rm rel} \sim \frac{D_0}{s_1s_2}.
\label{eq:collective_rate}
\end{equation}
Equivalently, this scaling corresponds to the geometric mean of the two
local rates, $\Gamma_{\rm rel} = \sqrt{\Gamma_1\Gamma_2} $, 
which should be understood as a scaling estimate for
the lowest collective mode, rather than as an additional independent
relaxation rate or an exact expression for the eigenvalue.

Equation~\eqref{eq:collective_rate} gives

$\tau_{\rm rel} \sim \frac{1}{\Gamma_{\rm rel}} \sim \frac{s_1s_2}{D_0}$.

Using Eq.~\eqref{eq:s1s2}, the corresponding characteristic frequency
therefore scales as

\begin{equation}
\nu \sim \frac{D_0}{\lambda^2} \frac{1}{1-\delta^2}.
\label{eq:nu_scaling_discussion}
\end{equation}
where numerical factors and the dependence on $\Delta$ are absorbed into
the reference frequency $\nu_0(\Delta)$.

This has the same asymmetry dependence as the empirical form Eq.\eqref{eq:dimensional_resonance} obtained from
the numerical results with $b_0$ close to unity, indicating that the geometrical
scaling captures the dominant dependence on the asymmetry.

This interpretation also clarifies why the simpler mean-field estimate
fails. Treating the ON-state localization and OFF-state diffusion as two
independent sequential processes gives
\begin{equation}
\tau_{\rm rel}^{\rm naive} \sim \tau_{\rm on}+\tau_{\rm off}.
\end{equation}
Such an additive construction neglects the coupling of the two spatial
sectors through the Fokker--Planck matching conditions and therefore does
not capture the collective nature of the relaxation mode.

However, a common argument
\cite{Ajdari1992,Julicher1995,julicher1997modeling,Rousselet1994,Xu_2021} posits that
the particle current can reach its maximum when the external driving
frequency is commensurate with the two characteristic timescales: the
relaxation time to the potential minimum in the ‘on’ state,
$\tau_{\rm on} = \frac{(1+|\delta|)^2}{4 \Delta}$, 
and the diffusive time to the adjacent potential maximum in the ‘off’ state,
$  \tau_{\rm off} = \frac{(1-|\delta|)^2}{8}$. 
This leads to a mean-field estimate for the resonance frequency, 
\begin{align}
      \label{eq:tau_mf}
\nu &= \left( \tau_{\rm on} + \tau_{\rm off} \right)^{-1} \\ \nonumber
&= \frac{8 \Delta}{(\Delta+2)(1+\delta^2)+(4-2\Delta)|\delta|}.
\end{align}

This predicts a decrease of the resonance frequency with increasing
$|\delta|$, in direct contradiction to the numerical results. The difference
between the two estimates therefore has a simple physical origin: the
naive expression treats the two parts of the relaxation as independent
sequential processes, whereas the Fokker--Planck problem describes a
single relaxation mode extending over the complete asymmetric unit cell.

The argument above should be regarded as a scaling interpretation rather
than an exact evaluation of the lowest eigenvalue of the piecewise
Fokker--Planck operator. Nevertheless, it identifies the geometrical
combination $s_1s_2$ that controls the leading asymmetry dependence and
provides a natural explanation for the observed $1-\delta^2$ structure.

\subsection{Implications and concluding remarks}

The results presented above provide a compact description of the transport
properties of a flashing linear asymmetric ratchet in terms of three
parameters: the spatial period $\lambda$, the asymmetry $\delta$, and the
barrier height $\Delta$. The particle current is characterized by a
well-defined resonance as a function of the switching frequency. Within
the regime of weak to moderate asymmetry, the current can be approximately
separated into an amplitude and a frequency-dependent part,
$
\langle\tilde{j}\rangle \simeq \widetilde{J}(\delta,\Delta) \frac{\nu f}{\nu^2+3\nu f+f^2},
$
where the amplitude is predominantly linear in $\delta$,
$
\widetilde{J}(\delta,\Delta) \simeq v_0(\Delta)\delta,
$
while the characteristic frequency is described by
$
\nu(\delta,\Delta) \simeq \frac{\nu_0(\Delta)} {1-b_0\,\delta^2}.
$
Together, these relations provide a simple phenomenological description of
the current over the range of parameters for which the numerical scaling
collapse is observed. The use of $\delta$ as the asymmetry parameter is
particularly convenient because it changes sign under spatial inversion,
making the odd symmetry of the current and the even symmetry of the
resonance frequency explicit.

An important consequence of the analysis is that the resonance cannot be
understood solely in terms of the sum of an ON-state localization time and
an OFF-state diffusion time. Such a description treats the two processes as independent sequential
events and yields a resonance frequency that decreases with increasing $|\delta|$, in direct contrast to the
observed increase of the resonance frequency with $|\delta|$.
The Fokker--Planck structure instead suggests that the
relevant relaxation is a collective mode of the two branches of the
asymmetric unit cell. The resulting scaling with $s_1s_2$ naturally
produces the observed $(1-\delta^2)^{-1}$ dependence. This provides a geometrical
interpretation of the resonance that does not rely on fitting the
relaxation time to a sum of independently estimated timescales.

The present analysis is deliberately restricted to the linear asymmetric
potential and to symmetric stochastic switching. The numerical agreement
between the Fokker--Planck calculation and Brownian dynamics indicates that
the scaling relations are robust within this setting, but it remains to be
seen how the same structure is modified for more general potential shapes,
non-symmetric switching protocols, or additional nonequilibrium driving.
In particular, the scaling argument based on the coupled spatial modes
suggests that the relevant geometrical combinations of the different
segments of a potential may provide a useful starting point for such
generalizations.

In summary, we have shown that a flashing linear asymmetric ratchet
exhibits a tunable resonance in its directed current and that both the
current amplitude and the resonance frequency admit simple scaling forms
in terms of the potential asymmetry and barrier height. The quadratic
dependence of the resonance frequency on the natural asymmetry parameter
and its connection to the product of the two branch lengths provide a
simple geometrical interpretation of the frequency selection mechanism.
These results offer a compact description of the transport response of the
flashing ratchet and may be useful for comparing different realizations of
periodically switched asymmetric potentials.

\section*{Acknowledgements}
Dipanjan Chakraborty acknowledges financial support from DST-SERB
through grant No. CRG/2021/003640. Debasish Chaudhuri acknowledges
financial support from the Department of Atomic Energy (DAE) through
Grant No. 1603/2/2020/IoP/R\&D-II/15028, a Visiting Professorship at
CY Cergy Paris Universit{\'e}, and an Associateship of IIT Bombay.

% \newpage   

% \newpage 
	    
% %\onecolumngrid

\appendix

 \section{Fokker-Planck equation} 
 \label{sec_appA}
 
%========================
The coupled Fokker--Planck equations for probability distributions  in the main text  expressed in terms of the dimensionless
variables, can be written explicitly as
\begin{multline}
\label{eq:fokker_planck_explicit_scaled1}
\partial_t P_{\rm on}^{\alpha}-\partial_{x}^2 P_{\rm on}^{\alpha}-F_\alpha \partial_xP_{\rm on}^{\alpha}(x,t)
= -fP_{\rm on}^{\alpha} +fP_{\rm off}^{\alpha},
\\
\partial_t P_{\rm off}^{\alpha} -\partial_{x}^2 P_{\rm off}^{\alpha}
= -fP_{\rm off}^{\alpha}+fP_{\rm on}^{\alpha},
\end{multline}
Here, the superscript $\alpha=1,2$ in the space- and time-dependent probability distributions $P_{\rm on}^{\alpha}(x,t)$ and $P_{\rm off}^{\alpha}(x,t)$ denotes the two spatial regions, $0\leq x\leq s_1/\lambda$ and $s_1/\lambda\leq x\leq 1$, respectively.
The piecewise-constant forces in the on state are given by
\begin{equation}
\label{eq:piecewise_forces}
F_1=\frac{2\Delta}{1-\delta},
\qquad
F_2=-\frac{2\Delta}{1+\delta}.
\end{equation}

To obtain the stationary solution of
Eq.~\eqref{eq:fokker_planck_explicit_scaled1}, we set
$\partial_t P=0$ and introduce, in each spatial region, the
two-component probability vector
\begin{equation}
\label{eq:probability_vector}
\boldsymbol{P}^{(\alpha)}(x)
=
\begin{pmatrix}
P_{\rm on}^{(\alpha)}(x)\\
P_{\rm off}^{(\alpha)}(x)
\end{pmatrix}.
\end{equation}
The stationary coupled Fokker--Planck equations can then be written
compactly as
\begin{equation}
\label{eq:stationary_fp_operator}
\mathcal{L}_{\alpha} \boldsymbol{P}^{(\alpha)}(x)=0,
\end{equation}
where
\begin{equation}
\label{eq:fp_operator_matrix}
\mathcal{L}_{\alpha} =
\begin{pmatrix}
\partial_x^2+F_\alpha\partial_x-f & f\\
f & \partial_x^2-f
\end{pmatrix}.
\end{equation}
The probability distributions can be expressed in termes of the complete set of spatial eigenmodes of the
form
\begin{equation}
\label{eq:spatial_eigenmode}
\boldsymbol{\phi}_n^{(\alpha)}(x)
=
\boldsymbol{u}_n^{(\alpha)}
e^{\Lambda_n^{(\alpha)}x}.
\end{equation}
Substitution of \eqref{eq:spatial_eigenmode} into
\eqref{eq:stationary_fp_operator} gives
\begin{equation}
\label{eq:eigenvalue_equation}
\boldsymbol{D}_{\alpha}
\left(\Lambda_n^{(\alpha)}\right)
\boldsymbol{u}_n^{(\alpha)}
=0,
\end{equation}
where
\begin{equation}
\label{eq:mat_D}
\boldsymbol{D}_{\alpha}(\Lambda)
=
\begin{pmatrix}
\Lambda^2+F_\alpha\Lambda-f & f\\
f & \Lambda^2-f
\end{pmatrix}.
\end{equation}
A nontrivial solution of \eqref{eq:eigenvalue_equation} requires det$[{\bf D}_\alpha]=0$
giving the characteristic equation 
\begin{equation}
\label{eq:characteristic_lambda}
\Lambda
\left[
\Lambda^3
+F_\alpha\Lambda^2
-2f\Lambda
-fF_\alpha
\right]
=0.
\end{equation}
Thus, in each spatial region there are four spatial eigenvalues,
$\Lambda_n^{(\alpha)}$, with $n=1,\ldots,4$, one of which is
\begin{equation}
\label{eq:zero_eigenvalue}
\Lambda_1^{(\alpha)}=0.
\end{equation}

For each eigenvalue $\Lambda_n^{(\alpha)}$, the corresponding
eigenvector can be chosen as
\begin{equation}
\label{eq:eigenvector}
\boldsymbol{u}_n^{(\alpha)}
=
\begin{pmatrix}
1\\
r_n^{(\alpha)}
\end{pmatrix}.
\end{equation}
Substitution of \eqref{eq:eigenvector} into \eqref{eq:eigenvalue_equation} gives
\begin{equation}
\label{eq:eigenvector_ratio}
r_n^{(\alpha)}
=
-\frac{f}
{\left(\Lambda_n^{(\alpha)}\right)^2-f}.
\end{equation}
For the zero mode in \eqref{eq:zero_eigenvalue},
\eqref{eq:eigenvector_ratio} gives
$r_1^{(\alpha)}=1$. Thus, the zero eigenvalue corresponds to a
spatially uniform mode with equal contributions to the on and off
probability densities.

The stationary probability distribution in each spatial region can
therefore be expressed as an eigenfunction expansion,
\begin{equation}
\label{eq:eigenfunction_expansion}
\boldsymbol{P}^{(\alpha)}(x)
=
\sum_{n=1}^{4}
C_n^{(\alpha)}
\boldsymbol{u}_n^{(\alpha)}
e^{\Lambda_n^{(\alpha)}x}.
\end{equation}
Using \eqref{eq:eigenvector}, the two components of
\eqref{eq:eigenfunction_expansion} are explicitly
\begin{align}
\label{eq:pon_eigen_expansion}
P_{\rm on}^{(\alpha)}(x)
&=
\sum_{n=1}^{4}
C_n^{(\alpha)}
e^{\Lambda_n^{(\alpha)}x},
\\
\label{eq:poff_eigen_expansion}
P_{\rm off}^{(\alpha)}(x)
&=
\sum_{n=1}^{4}
C_n^{(\alpha)}
r_n^{(\alpha)}
e^{\Lambda_n^{(\alpha)}x}.
\end{align}
Thus, the relative amplitudes of the on and off components of each
spatial eigenmode are fixed by the corresponding eigenvector through
\eqref{eq:eigenvector_ratio}. Consequently, there are four
independent expansion coefficients in each spatial region, giving a
total of eight unknown coefficients
$\{C_n^{(1)},C_n^{(2)}\}$.

The coefficients are determined by requiring continuity of the
probability densities and probability currents at $x=s_1/\lambda$, together
with periodicity across the boundaries $x=0$ and $x=1$. For the
Fokker--Planck equations in
\eqref{eq:fokker_planck_explicit_scaled1}, the probability currents in
the on and off states are
\begin{align}
\label{eq:current_on}
J_{\rm on}^{(\alpha)}(x)
&=
-\partial_xP_{\rm on}^{(\alpha)}(x)
-F_\alpha P_{\rm on}^{(\alpha)}(x),
\\
\label{eq:current_off}
J_{\rm off}^{(\alpha)}(x)
&=
-\partial_xP_{\rm off}^{(\alpha)}(x).
\end{align}
Introducing the current vector
\begin{equation}
\label{eq:current_vector}
\boldsymbol{J}^{(\alpha)}(x)
=
\begin{pmatrix}
J_{\rm on}^{(\alpha)}(x)\\
J_{\rm off}^{(\alpha)}(x)
\end{pmatrix},
\end{equation}
the continuity conditions at 
$x=s_1/\lambda$, are
\begin{equation}
\label{eq:matching_x0}
\boldsymbol{P}^{(1)}(s_1/\lambda)
=
\boldsymbol{P}^{(2)}(s_1/\lambda),
\qquad
\boldsymbol{J}^{(1)}(s_1/\lambda)
=
\boldsymbol{J}^{(2)}(s_1/\lambda).
\end{equation}
Similarly, periodicity of the probability densities and currents over
one spatial period requires
\begin{equation}
\label{eq:matching_periodic}
\boldsymbol{P}^{(1)}(0)
=
\boldsymbol{P}^{(2)}(1),
\qquad
\boldsymbol{J}^{(1)}(0)
=
\boldsymbol{J}^{(2)}(1).
\end{equation}

Equations \eqref{eq:matching_x0}, and \eqref{eq:matching_periodic} provide the
matching conditions for the eight expansion coefficients appearing
in \eqref{eq:eigenfunction_expansion}. Owing to conservation of total
probability, one of these matching conditions is linearly dependent,
so that the stationary solution is determined only up to an overall
multiplicative constant. This constant is fixed by imposing the
normalization condition
\begin{equation}
\label{eq:probability_normalization}
\int_0^1
\mathrm{d}x\,
\left[
P_{\rm on}(x)+P_{\rm off}(x)
\right]
=1.
\end{equation}
The eigenvalues obtained from \eqref{eq:characteristic_lambda},
together with the matching conditions in \eqref{eq:matching_x0} and
\eqref{eq:matching_periodic} and the normalization condition in
\eqref{eq:probability_normalization}, uniquely determine the
stationary probability distributions. The resulting system of
algebraic equations for the expansion coefficients is then solved
numerically.

\section{Dependence of current on the barrier height}
\label{sec:current_on_Delta}

\begin{figure}[!t]
  \centering
  \includegraphics[width=0.5\linewidth]{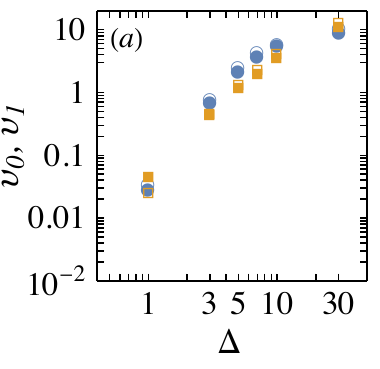}
  \caption{Dependence of the weak-asymmetry coefficients $v_0$ and
    $v_1$ of the dimensionless current amplitude
    $\tilde{\mathcal{J}}(\delta,\Delta)$ on $\Delta$. Circles
    (squares) denote $v_0$ ($v_1$); open (filled) symbols show
    Fokker--Planck (Brownian-dynamics) results. Both coefficients
    increase with $\Delta$ and remain comparable in magnitude, with
    $v_0$ tending to saturate at large $\Delta$ while $v_1$ continues
    to increase.}
  \label{fig:current_amplitude_coeffs}
\end{figure}

Figure~\ref{fig:current_amplitude_coeffs} shows the dependence of the
coefficients $v_0$ and $v_1$ in the weak-asymmetry expansion of the
dimensionless current amplitude $\tilde{\mathcal{J}}(\delta,\Delta)$
on the potential height $\Delta$. The results obtained from the
Fokker--Planck equations are in good agreement with the
Brownian-dynamics simulations over the range considered. Both
coefficients increase with $\Delta$ and remain of comparable
magnitude. While $v_0$ tends to saturate at large $\Delta$, $v_1$
continues to increase, indicating that the nonlinear correction
becomes progressively more relevant at larger barrier heights.  For
sufficiently weak asymmetry, however, the leading-order dependence
remains linear in $\delta$, so that
\begin{equation}
\tilde{\mathcal{J}}(\delta,\Delta)\simeq v_0(\Delta)\delta.
\label{eq:amplitude_linear}
\end{equation}
For example, at $\Delta=1$, a linear fit gives $v_0\simeq0.025$.

\begin{figure}[!ht]
  \centering
  \includegraphics[width=\linewidth]{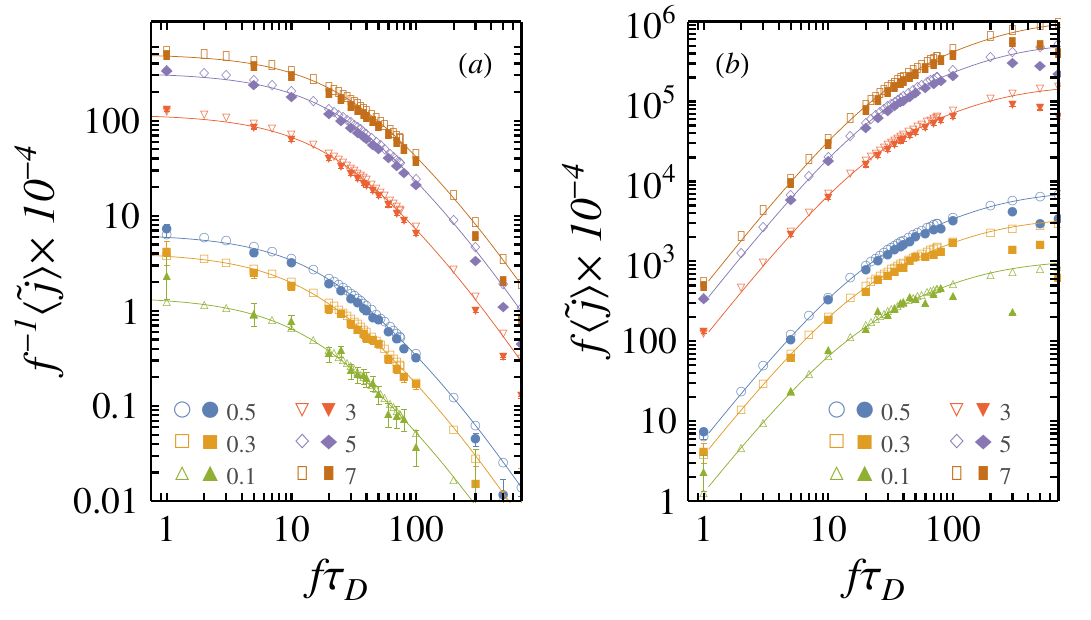}    
  \caption{\textbf{(a)} $f^{-1}\jtilde$ versus $f$ for different
    values of $\dl$ and $\D$. Circles, squares, and upright triangles
    correspond to varying $\dl$ at $\D=1$, while inverted triangles,
    diamonds, and rectangles correspond to varying $\D$ at $\dl=0.5$,
    as indicated in the legends. \textbf{(b)} $f\jtilde$ versus $f$
    for the same parameter sets. In both panels, open and filled
    symbols denote results from the Fokker--Planck equation and
    Brownian dynamics simulations, respectively.}
  \label{fig:high_low_freq}
\end{figure}

\section{High and Low frequency limits}
\label{app:high_low_frequency}

Figure~\ref{fig:high_low_freq} illustrates the low- and high-frequency
scaling of $\jtilde$ for different values of the parameters $\dl$ and
$\D$. In Fig.~\ref{fig:high_low_freq}(a), plotting $f^{-1}\jtilde$
versus $f$ shows that the curves approach a constant as $f\to0$. This
behavior confirms the linear low-frequency scaling, $\jtilde\sim
f$. In contrast, Fig.~\ref{fig:high_low_freq}(b) shows that $f\jtilde$
approaches a constant as $f\to\infty$, demonstrating the
inverse-frequency scaling $\jtilde\sim f^{-1}$ at high frequencies.
The scaling behavior is robust across the range of $\dl$ and $\D$
considered. The open symbols represent numerical solutions of the
Fokker--Planck equation, while the filled symbols correspond to
Brownian dynamics simulations. The close agreement between the two
sets of results provides a numerical validation of the predicted
frequency dependence.

\bibliographystyle{unsrt}
\bibliography{paper,ratchet_2d}

%\end{appendices}
\end{document}